\documentclass[doublecol]{epl2}

\usepackage{amssymb,amsmath,latexsym}

\newcommand\beq{\begin{equation}}
\newcommand\eeq{\end{equation}}

\newcommand\bea{\begin{eqnarray}}
\newcommand\eea{\end{eqnarray}}

\newcommand\bi{\begin{itemize}}
\newcommand\ei{\end{itemize}}

\newcommand\non{\nonumber}

\newcommand\tstub{\mathcal T}

\newcommand\ie{{\textsl{i.e.}}}

\newcommand\scurrent{{\textsf{SC~}}}
\newcommand\scurrentd{{\textsf{SC}}}

\newcommand\nsnd{{\textsf{NSN}}}

\newcommand\card{{\textsf{CAR}}}
\newcommand\ard{{\textsf{AR}}}

\newcommand\nsn{{\textsf{NSN~}}}

\newcommand\car{{\textsf{CAR~}}}
\newcommand\ar{{\textsf{AR~}}}

\def\dfrac#1#2{{\displaystyle\frac{#1}{#2}}}

\newif\ifboo \boofalse
\def\Review#1{\boofalse{\it #1},}

\def\Vol#1{\ifboo Vol. {\bf #1}\else{\bf #1}\fi}
\def\Year#1{\ifboo #1\else(#1)\fi}

\def\Page#1{\ifboo {\rm p. #1}\else{\rm #1}\fi}

\revision{\title{Resonant charge and spin transport in a
$\tstub$-stub coupled to a superconductor}}
\revision{
\shorttitle{Resonant charge and spin transport through a $\tstub$-stub}}

\author{Sourin Das\inst{1,2} \and Sumathi Rao\inst{3} \and Arijit
Saha\inst{3}}
\shortauthor{Sourin Das \etal}

\institute{
\inst{1} {Institut f{\"u}r Festk{\"o}rper-Forschung--Theorie3,
Forschungszentrum J{\"u}lich, 52425 J{\"u}lich, Germany}\\
\inst{2} {Institut f{\"u}r Theoretische Physik A, RWTH Aachen,
52056 Aachen, Germany}\\
\inst{3} {Harish$-$Chandra  Research Institute, Chhatnag Road, Jhusi,
Allahabad 211 019, India}}


\abstract{We study transport through a single channel $\tstub$-stub
geometry strongly coupled to a superconducting reservoir. In contrast to
the standard stub geometry which has both transmission resonances and
anti-resonances in the coherent limit, we find that due to the proximity
effect, this geometry shows neither a $T=1$ resonance ($T$ is the 
transmission probability for electrons incident on the $\tstub$-stub) nor 
a $T=0$ anti-resonance as we vary the energy of the incident electron.
Instead, we find that there is only one resonant value at $T=1/4$, where
charge transport vanishes while the spin transport is perfect.}

\begin{document}

\maketitle

\section{Introduction}
One of the many intriguing issues in the subject of
spintronics~\cite{review} concerns production and detection of pure spin
current. A simple minded but popular proposal for production of pure spin
current (\scurrentd) involves electrons flowing with equal flux in opposite
directions with opposite spin polarizations. This situation results in the
exact cancellation of the charge current while the spin current adds up.
Alternatively one can also produce pure \scurrent by having a
unidirectional flow of electrons and holes together with equal flux and
spin polarization. In this case also, the charge current cancels out
leaving behind a pure spin current. In this paper we will work along the
lines of the second proposal for production of pure \scurrent.

Situations involving current carried by an admixture of electrons and
holes are naturally realized in systems involving superconductor-normal
junctions~\cite{andreev,blonder,hekking1,fazioreview,russo,belzig2008,
chandrasekhar,zaikin2007} due to the interplay of Andreev reflection and
normal reflection at the interface between the superconductor and the
normal metal. A recent proposal by the present authors which exploited this
fact for production of pure \scurrent involved transport of electrons and
holes across a normal-superconducting-normal (\nsnd) junction
~\cite{das2007drsahaepl}. The normal metals in the \nsn junction
were considered to be one-dimensional interacting electron gases 
which were modelled as Luttinger liquid wires, and a situation
corresponding to pure spin current was shown to be an unstable fixed
point of the theory. In contrast to our previous study, here we work with
free electrons and pure spin current is produced due to resonant
conversion of incident electrons into transmitted electrons or holes with
equal amplitude across a $\tstub$-stub geometry which is strongly coupled
to a superconducting reservoir. Inclusion of electron-electron interaction
can lead to very interesting physics~\cite{nazarov} in the
presence of resonances but this is beyond the scope of the present work.

\begin{figure}[htb]
\begin{center}
\includegraphics[width=5cm,height=3.5cm]{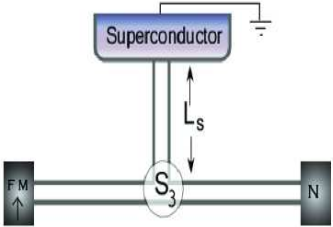}
\hskip 0cm \caption{Cartoon of the set-up proposed for the
$\tstub$-stub geometry. $L_{s}$ is the length of the stub and
$\mathbb{S}_3$ represents the $\mathbb{S}$-matrix describing the
three-wire junction.}
\label{stub}
\end{center}
\end{figure}

Theoretically, although our model appears simple, it is one
of the first models to realize resonant
transmission of electrons through a complex barrier (in this case a stub,
which hosts both electron and hole waves due to its coupling to
superconductor at one end) with an amplitude which is not unimodular.
To develop an
understanding of the new resonance, we
explore the analytic structure of the electron transmission amplitude in
the complex energy plane. We show that the analytic structure of the
transmission function is quite different from the cases of standard double
barrier resonances or the resonance-anti resonance pairs of the normal stub
geometry~\cite{porod,takane}.
In this article, we consider a single mode $\tstub$-stub which is
coupled to a superconducting reservoir at one end, and to a single mode wire
at the other end,  which is then connected to the left and right reservoirs
(see Fig.\ref{stub}). We assume that the current injected from  the left
reservoir into the wire is completely spin polarized. Additionally we
assume that the bias which drives current between the left and the right
reservoir is such that the net current is always flowing from left to
right. Even though our calculations use a single mode approximation,
these results can be applied to a wire with multiple channels as long as
the scattering matrix at the junction does not mix the channels
substantially at the junction - $i.e.$, as long as the $S$-matrix is block
diagonal, in each of the channels, or can be approximated as being almost
block diagonal. In the general case, with substantial inter-mode scattering
at the junction, detailed numerical analysis, beyond the scope of this
analysis, would be required to compute the transmission function.

Here the role of the bulk superconductor is to convert an incoming
electron in the stub into an outgoing hole via Andreev
reflection (\ard). We restrict ourselves to the case where the coupling
between the superconducting reservoir and the stub is perfect, so that the
\ar probability of an incident electron to reflect as a hole is unity.

We find that as we vary the energy of the incident electrons for fixed 
length of the stub or the length of the stub for fixed incident energy, due
to the proximity effect we get a resonance at $T=1/4$ where the charge
current across the stub is identically zero while spin current adds up.
Thus, we get resonant transport of pure \scurrentd, which is exactly
matched by the ``anti-resonant'' transport of the total charge current. The
advantage of this set-up is that the detection of the zero total charge
current across the $\tstub$-stub automatically ensures simultaneous
detection of pure spin current. Note that because of the perfect conversion
of electrons to holes (and vice-versa) at the boundary of the stub and the
superconductor, the scattering matrix at the three wire junction
(describing transport between lead-$1$ and lead-$2$, see Fig.~\ref{stub})
can be parameterized by an effective $4\times4$ $\mathbb{S}$-matrix. This
includes both the particle and the hole sector. Transport across the
junction involves two new processes other than the usual reflection and
transmission of electrons (holes) - {\sl{(a)}} transmission of an incident
electron state into a hole state across the stub,~which is
usually called crossed Andreev reflection (\card) and {\sl{(b)}}
reflection of an incident electron state into a hole state from the stub
(defined earlier as \ard). These processes result from the fact
that an electron incident on the junction can tunnel into the stub branch
and bounce back to the junction as a hole. In fact, it can bounce back and
forth between the superconductor and the three wire junction undergoing
multiple electron-hole conversions. If the incident electron undergoes an
even number of bounces, it will come out as an electron but if it undergoes
an odd number of bounces, it will come out as a hole. Hence when the
electron exits the stub after an odd number of bounces, it contributes to
the electron-hole amplitudes. If it exits the stub branch as a hole on the
same wire, from where it entered, it will contribute to the \ar and if it
exits on the other wire, it will contribute to the \card. Hence, even
though the wires-$1$,$2$ (coupled to the reservoir-$1$,$2$, see
Fig.~\ref{stub}) are not directly coupled to the superconductor and there
are no Andreev processes at the three wire junction, the stub allows for
finite amplitudes for processes which effectively mimic \car process
between wire-$1$ and wire-$2$. At low temperatures ($L_{th} > L_s$, where
$L_{th}$ is the thermal length and $L_s$ is the stub length) the
stub branch acts like a coherent particle-hole resonator leading
to resonances in the amplitude for the effective \car process and the
direct electron transmission process, due to quantum interference between
the particle and hole waves inside the stub. {\it The resonant production of
electrons and holes in wire-$2$ in response to incident electrons from
wire-$1$ on the $\tstub$-stub, due to quantum interference between
particle and hole waves inside the stub, can lead to resonant
production of pure \scurrent in wire-$2$. This is the main result of this
paper.}

\section{Quantum mechanics of the $\tstub$-stub}
We start with a junction of three wires, with two of them connected to
a fully spin polarized reservoir and a normal reservoir respectively (see
Fig.\ref{stub}) and the third one (a stub of finite length $L_{s}$)
coupled to a superconducting reservoir. A normal stub would have boundary
conditions at the stub end, which would completely reflect an incoming
electron to an outgoing electron. Here, on the other hand, the
superconducting reservoir turns an incoming electron completely to an
outgoing hole in the perfect Andreev limit. The $3\times3$
$\mathbb{S}$-matrix coupling the wires to the stub initially is a `normal'
scattering matrix for electrons or holes with spin-up ($\uparrow$) or
spin-down ($\downarrow$) and left-right symmetry. We have considered the
$\mathbb{S}$-matrix to be the same for the electron and hole sector
assuming electron-hole symmetry.
\begin{figure*}[htb]
\begin{center}
\vskip +1.0cm
\includegraphics[width=15cm,height=6cm]{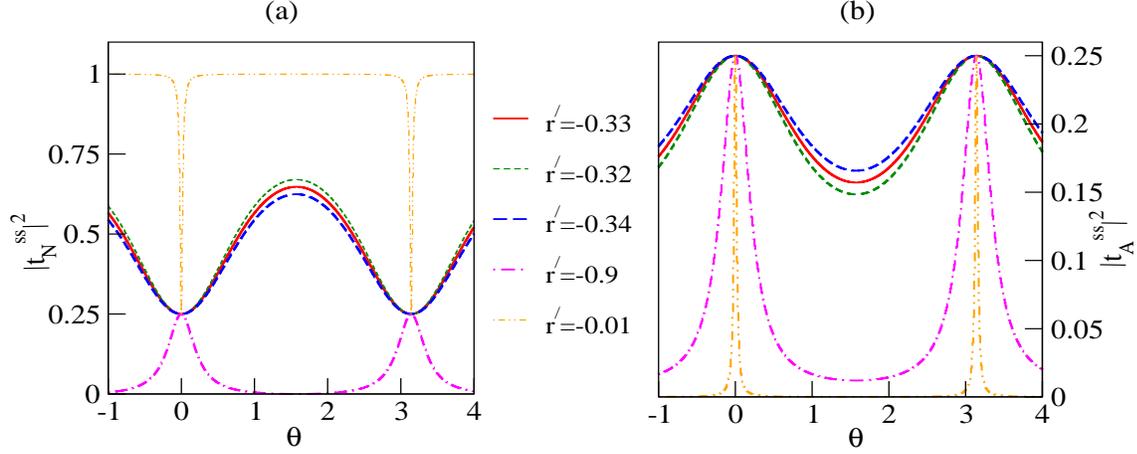}
\hskip 0.0cm \caption{(Color online)(a) Transmission probability
$|t_{N}^{ss}|^{2}$ and (b) \car probability $|t_{A}^{ss}|^{2}$
are plotted in units of $e^{2}/h$ as a function of $\theta=2EL_{s}/\hbar
v_{F}-\cos^{-1}(E/\Delta)$ for five different values of $r^{\prime}$.}
\label{superstub1}
\end{center}
\end{figure*}
As studied earlier~\cite{das2004drs}, the $\mathbb{S}$-matrix
for a three-wire junction can be described by a single parameter, if
we choose real parametrization. In terms of the single parameter
$r^\prime$, the $\mathbb{S}$-matrix describing electrons
($\uparrow/\downarrow$) or holes ($\uparrow/\downarrow$) is given by
\beq
\mathbb{S}_{\mathrm{3}} \quad = \quad \begin{bmatrix} ~r^{\prime} &
t^{\prime} & t ~\\
~t^{\prime} & r^{\prime} & t ~\\
~t & t & r ~ \label{S3}
\end{bmatrix} ~,
\eeq
where using unitarity, we have
 \beq
t^\prime=1+r^\prime,
r=-1-2r^\prime, \\
t=\sqrt{(-2r^\prime)(1+r^\prime)}~, \label{smatx1}
 \eeq
and $-1\leqslant r^{\prime}\leqslant0$.
Here, we consider the single parameter $r^{\prime}$ to denote the
degreee of coupling between the wires $1$ and $2$ and the stub. For
$r^{\prime}=-1$, both the wires and the stub are all disconnected from each
other. For $r^{\prime}=0$, the two wires are fully connected to one
another, but disconnected from the stub. At $r^{\prime}=-0.5$, $r=0$,
which means that multiple reflections within the stub do not occur,
although the wire is well-coupled to the stub. Another interesting value
for $r^{\prime}$ is $r^{\prime}=-1/3$ as this corresponds to the most
symmetric $\mathbb{S}$-matrix describing the junction. The other end of 
the stub  is assumed to be perfectly coupled to a superconducting
reservoir, and the $2\times 2$ $\mathbb{S}$-matrix describing the
superconducting boundary is given by
\beq
\mathbb{S}_{\mathrm{bound}} \quad = \quad \begin{bmatrix} ~r^{s} &
r_{a}^{s} ~\\
~r_{a}^{s} & r^{s} ~
\end{bmatrix}~,
\label{SA} \eeq
which, in the perfect Andreev limit~\cite{beenaker} has $r_{a}^{s}=\alpha
= e^{-i \cos^{-1}(E/\Delta)} \times e^{\pm i \phi}$ and $r^{s}=0$ .
For $E=0$, $\alpha$ reduces to $-i e^{\pm i \phi}$ where $\phi$ is an
energy independent part. When the \ar is not perfect, \ie~$r^s
\ne 0$.

Now, let us compute the effective $4\times 4$ $\mathbb{S}$-matrix
coupling wire-$1$ and wire-$2$. As discussed in the introduction,
since the stub hosts both electrons and holes, the junction of wire-$1$
and wire-$2$ effectively has a $\mathbb{S}$-matrix very similar to a
normal-superconducting-normal (\nsnd) junction. The coherent amplitudes
for the reflection, transmission, \ar and \car can be calculated by adding
up all possible Feynman paths taken by the incident electron while
bouncing back and forth inside the $\tstub$-stub. These amplitudes depend
only on a single parameter $r^\prime$ which parameterize the $3\times 3$
$\mathbb{S}$-matrix at the junction and are given by
\bea r_{N}^{{ss}} &=&
r^{\prime}+\dfrac{\alpha^{2}rt^{2}\eta^{2}}{(1-\alpha^{2}r^{2}\eta^{2})}
\label{rN} ~,
\\
t_{ N}^{{ss}} &=&
t^{\prime}+\dfrac{\alpha^{2}rt^{2}\eta^{2}}{(1-\alpha^{2}r^{2}\eta^{2})}
\label{tN}
 ~,
\\
 r_{A}^{{ss}} &=& t_{A}^{ss}=\dfrac {\alpha t^{2} \eta e^{i
\phi}}{(1-\alpha^{2}r^{2}\eta^{2})}\label{tA} ~, \eea
where $r_{N}^{ss}$, $t_{N}^{ss}$, $r_{A}^{ss}$, $t_{A}^{ss}$ are the
reflection, transmission, \ar and \car amplitudes respectively. The
superscript $ss$ stands for the case of superconducting stub where
the stub is strongly coupled to the superconductor while the subscript
$N$ or $A$ corresponds to normal or Andreev processes. The $r,t$ and
$t^{\prime}$ were defined earlier in terms of $r^\prime$ in
Eq.~\ref{smatx1}. Note the fact that the amplitudes for the \ar and \car
are the same owing to the left-right symmetry of the $S$-matrix representing
the junction. The  energy dependent phase $\eta$ comes
from adding the paths of the incident electron, which gets converted
to a hole at the superconducting boundary, at the first bounce and all odd
bounces and back to an electron at all even bounces. Here  $\eta =
e^{i2EL_{s}/\hbar v_{F}}$ and  $E$ is the energy of the incident electron
with respect to Fermi energy in the superconductor ($E_F$) and
$L_{s}$ is the length of the stub. For $E=0$ the phase $\eta$
of the propagating electrons and holes inside the stub cancel each other
as a hole follows exactly the time reversed path of an electron and
$\alpha$ is fixed to be $=-i$. Let us define $e^{i \theta}= \eta \alpha =
e^{i(2EL_s/\hbar v_F-\cos^{-1}(E/\Delta))}$ \ie~$\theta=2EL_s/\hbar
v_{F}-\cos^{-1}(E/\Delta)$ to be an energy dependent phase parameter.
The constant part of the \ar phase, $\phi$ may be set to zero without loss
of generality, since it does not affect the probability.

The variation of $|t_{N}^{ss}|^{2}$ and $|t_{A}^{ss}|^{2}$ as a function
of $\theta$ for various values of the parameter $r^\prime$ is shown in
Fig.~\ref{superstub1}. Note that when we tune the energy such that $\theta$
=$n\pi$ ($n$ is $0$ or an integer), both the transmission probability
($|t_{N}^{ss}|^{2}$) and \car probability ($|t_{A}^{ss}|^{2}$), are exactly
$=1/4$ for any value of the parameter $r^\prime$ \ie, independent of the
details of the $\mathbb{S}_{3}$ matrix. This resembles the resonance $T=1$
(anti-resonance $T=0$) of a standard stub geometry. Also, note that the
\car probablity is maximum at the `resonant' value of $|t_{A}^{ss}|^{2}=1/4$
plotted as a function of $\theta$ (see Fig.~\ref{superstub1}(b)). On the
other hand, for the normal transmission $|t_{N}^{ss}|^{2}$ plotted as a
function of $\theta$, the `resonant value' of $1/4$ represents a maximum
for $r^\prime<-0.5$ and minimum for $r^\prime>-0.5$
(see Fig.~\ref{superstub1}(a)). This is a very peculiar feature of this
geometry. Hence $r^{\prime}=-1/2$ is the crossover point where the
transmission is completely flat as a function of the energy.  Note also
that the superconducting stub geometry does not host any true resonance
($|t_{N}^{ss}|^{2}$=1) or anti-resonance ($|t_{N}^{ss}|^{2}$=0) in sharp
contrast to a normal $\tstub$-stub or the double-barrier problem.

Now the general case is when the superconductor-stub
junction is not perfect and allows for both normal reflection and
\ard. In this case, an electron incident on the stub can enter the
stub and undergo either normal reflection at the superconductor-stub
junction with an amplitude $r^{s}$ or it can undergo \ar with an
amplitude $r_{a}^{s}$. Hence an incident electron from reservoir-$1$
can either reflect back as an electron in the same wire (wire-$1$)
or transmit as an electron into the opposite wire (wire-$2$) after
suffering a single reflection at the superconductor-stub junction
unlike the previous case where at least two bounces inside the stub
were required before an incident electron entering the stub could
reflect back into wire-$1$ as an electron or transmit into wire-$2$
as an electron. The coherent amplitudes for the transmission are
analytically very difficult to obtain from a Feynman path approach (as was
done earlier for the case of a perfect junction)~\ie, by summing
up the coherent amplitudes for all possible paths which take an electron
from the left to the right reservoir. However, for finite values of $r^s$,
we expect the new resonance to survive as the existence of this resonance 
just relies on the fact that there are two channels for transport (the 
electron and the hole channel) and the fact that they effectively mix 
with one another as far as the scattering matrix describing the stub is 
concerned. Hence for small values of $r^s$, the various transmission 
probabilities will not change substantially with respect to the case of 
$r^s=0$, close to the resonance.
\subsection{Pole structure analysis:-}
The subtle differences between the normal stub geometry and the
superconducting stub geometry emerge once we analyze the zeros and
the pole structure of the electron transmission amplitude for the two
cases. For the normal stub geometry, the position of zeros and the poles
of the electron transmission amplitude, $t_{N}^{ns}$ as a function of the
energy $E$ (parametrised as $\theta= 2EL_s/\hbar v_{F}$) in the complex
energy plane are given by
\bea \theta_Z^n &=& \theta_Z^{n0} + i 0\quad {\mathrm{(zeros)}}~, \non\\
\theta_P^n &=& \theta_P^{n0}  - \dfrac{i}{2} \ln \left(\dfrac{1}{2r^\prime +
1}\right) \quad {\mathrm{(poles)}}~, \eea
where $\theta_{Z/P}^n$ represents values of $E$ at the zeros and poles of
the transmission amplitude, $p$ is an integer and $\theta_{Z/P}^{n0}=p \pi$
represents the real part of the zero and the pole. Here the superscript $n$
stands for normal stub and the subscript $Z$ and $P$ stand for the
zeros and poles. In this case the zeros lie on the real energy axis,
which is why the transmission function has zeros, and the poles lie
in the complex plane off the real axis. But the real part of the poles
coincide with the zeros. This fact can be directly related to the symmetric
form of the transmission probability as a function of energy of the
incident electron~\cite{porod}. The stub has transmission maxima ($T^n =
|t_{N}^{ns}|^2=1$) at $\theta^{n}_m = (2p +1) \pi/2$ where the subscript
$m$ represents maximum transmission. Expanding the transmission probability
($|t_{N}^{ns}|^2$) around $\theta=\theta^{n}_m$, one obtains the following
expression,
\bea
T^n|_{\theta^{n}=\theta^{n}_m} &= & \dfrac{(2 \Gamma)^2}{(2\Gamma)^2+
(\theta^n-\theta^n_m)^2} + \dfrac{\Gamma^2 (\theta -
\theta^n_m)^2}{(2\Gamma)^2 +
(\theta^n-\theta^n_m)^2}    \non \\
&&\label{ns-lineshape}
\eea
where $\Gamma=t^\prime/r$. The first term on the right hand
side of Eq.~\ref{ns-lineshape} looks like the standard Lorentzian
observed for a double barrier and the presence of the second term leads to
a deviation from this shape which is expected for a Fano-type
resonance~\cite{porod}. But since here the real part of the zeros and poles
of the transmission amplitude coincide, we get a symmetric function in
energy around $\theta^n=\theta^n_m$. This is not true in general for
Fano-type resonances, which could have asymmetric line shapes. Also note
that the position of the transmission maxima ($T=1$, \ie, $\theta^{n}_m =
(2p +1) \pi/2$) on the real energy axis is exactly half way between the
position of the real part of the poles or the zeros
(\ie, $\theta_{Z/P}^{n0}=p \pi$).

Similarly, in case of the superconducting stub, we consider the
energy dependent zeros and poles of the electron transmission amplitude
$t_{N}^{ss}$ which are given by

\bea \theta_Z^s  &=& p \pi - \dfrac{i}{2} \ln
\left(\dfrac{1}{2r^\prime + 1}\right)
\quad {\mathrm{(zeros)}}, \non\\
\theta_P^s  &=& p \pi - i \ln
\left(\dfrac{1}{2r^\prime + 1}\right)
\quad {\mathrm{(poles)}}~, \eea
where $\theta_{Z/P}^s$ represents the values of $2EL_{s}/\hbar v_{F}-
\cos^{-1} (E/\Delta)$ at the zeros and poles and $p$ is an integer.

Analogous to the case of the normal stub described above, here too we find
that the real part of the zeros and the poles coincide with one another,
which ensures a symmetric line shape. But unlike the previous case here
the zeros do not lie on the real axis. This results in the absence of zeros
in the transmission probability for electrons across the superconducting
stub. Also, in this case there is no transmission unity
($T^s=|t^{ss}_N|^2=1$) unlike the case of the normal stub. The resonances
(maxima or minima depending on whether $r^\prime < -1/2$ or
$r^\prime > -1/2$, see Fig.~\ref{superstub1}) in $T^s$ appear at
$\theta^s_p = p \pi$. Expanding the $T^s$ around $\theta^s_p$, we
obtain the following line shape
\begin{figure*}[htb]
\begin{center}
\vskip +0.7cm
\includegraphics[width=15cm,height=6cm]{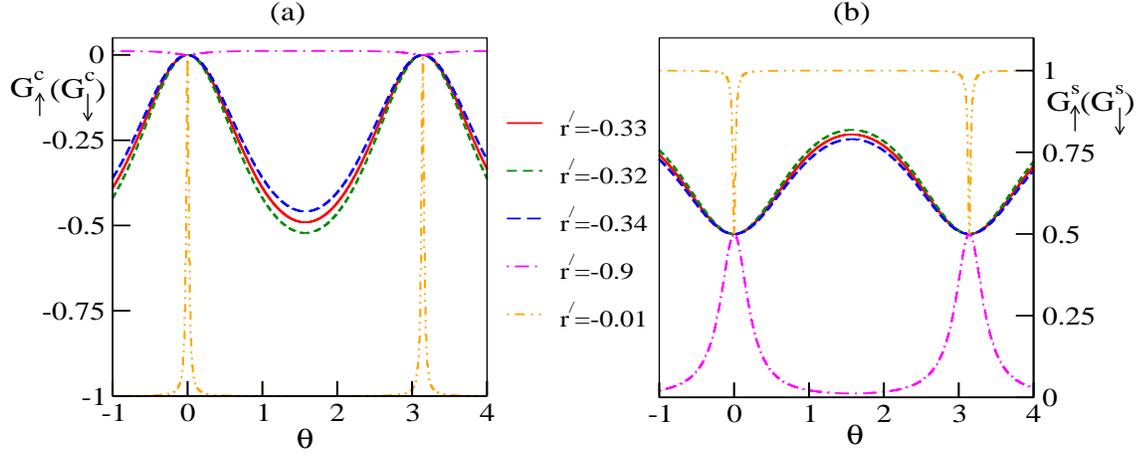}
\hskip 0cm \caption{(Color online)(a)~Charge conductance
$G_{\uparrow}^{c}$($G_{\downarrow}^{c}$) in units of $e^{2}/h$ and
(b)~Spin conductance $G_{\uparrow}^{s}$ $(G_{\downarrow}^{s})$ in units
of incident spin for spin $\uparrow$ ($\downarrow$) polarised electrons
plotted as a function of $\theta=2EL_{s}/\hbar v_{F}-\cos^{-1}(E/\Delta)$
for five different values of $r^{\prime}$.}\label{superstub2}
\end{center}
\end{figure*}
\bea T^s|_{\theta^s=\theta^s_p} &=& \dfrac{1}{4}\dfrac{(2 \Gamma_1
\Gamma_2)^2}{ (2\Gamma_1\Gamma_2) ^2 + (\theta^s-\theta_p^s)^2}
\non\\&&+ \dfrac{\Gamma_2^2 (\theta^s - \theta^s_p)^2}{
(2\Gamma_1\Gamma_2) ^2 + (\theta^s-\theta_p^s)^2}~,
\label{ss-lineshape} \eea
where $\Gamma_1=({2 r^\prime})/({1+2r^\prime})$ and $\Gamma_2 =
({1+r^\prime})/({1+2r^\prime})$.
As in the case of normal stub, the first term on the right hand side of
Eq.~\ref{ss-lineshape} represents a standard Lorentzian while
the presence of the second term leads to a deviation from it. But note
that the coefficient of the Lorentzian function which is unity for the
normal stub becomes one quarter for the superconducting stub;  this
results in a resonance in transmission probability which is not unity
but one quarter of unity. Also, if $r^{\prime}<-0.5$ the resonance
represents a transmission maxima for $T^s$ and if $r^{\prime}>-0.5$ the
resonance is a minima for $T^s$ as can be seen from the above equation
(whether it is a maximum or minimum depends on whether the zero or the
pole dominates). In this case also we get a perfectly symmetric line shape
(see Fig.~\ref{superstub1}) since the real part of the zeros and poles of
the transmission amplitude coincide with each other. Hence the main source
of difference between resonances for the superconducting and the normal
stub lies in the fact that for the superconducting stub the zeros of the
complex transmission amplitude lie in the complex plane off the real
$\theta$ axis while for the normal stub they lie on the real $\theta$ axis
itself. This explains the absence of zeros in the transmission amplitude
in the superconducting stub geometry.
\subsection{Charge and spin currents:-}
As an application of the above geometry, we note that this geometry can be
used to produce pure \scurrent. A similar proposal for the production of
pure \scurrent involving a \nsn junction was also discussed in the past
by the present authors in Ref.~\cite{das2007drsahaepl} as mentioned before.
In the present case, the pure \scurrent is shown to be coupled to the
new resonance discussed above. The main point to note here is that when an
electron is incident on the stub, then if the amplitudes for the normal
transmission $t^{ss}_N$ and \car $t^{ss}_A$ are identical, then the
probabilities for an incident electron to transmit as an electron or a
hole across the stub are identical, and effectively the transmitted charge
current will be zero. Furthermore, if the incident electron is spin
polarized, then the amplitudes for the transmitted electron ($t^{ss}_N$)
and holes ($t^{ss}_A$) will also have the same spin polarization as
long as the superconductor at the junction is a singlet superconductor.
This results in pure spin current with zero net charge current.

Thus, for spin polarized electrons when the superconducting stub is tuned
to resonance, \ie, $|t^{ss}_N|^2=|t^{ss}_A|^2=\frac{1}{4}$, the outcome will
be resonant production of pure \scurrent. In this resonant situation $25\%$
of the incident spin-up electrons get transmitted through the stub via the
normal transmission process and $25\%$ get converted to spin-up
holes via the \car process, as they pass through the $\tstub$-stub. Hence
the transmitted charge across the $\tstub$-stub is zero on the average,
but there is pure \scurrent flowing out of the system.
At zero temperature limit the linear response charge conductance
of incident spin polarized electrons is given by
$G_{\uparrow(\downarrow)}= e^2/h(|t^{ss}_A|^2 - |t^{ss}_N|^2)$ and
the spin conductance
$G^s_{\uparrow(\downarrow)} \propto (|t^{ss}_A|^2 + |t^{ss}_N|^2$).
These are depicted in Fig.\ref{superstub2}. The important point to
note in this geometry  is that the maxima in the \scurrent is accompanied
by zero charge current. At resonance, the charge of the incident electron
is absorbed by the superconductor, and no charge is either transmitted or
reflected from the $\tstub$-stub on the average. On the other hand, in units of
the original spin of the electron, probablistically on an average, half of
the spin is transmitted and the remaining half is reflected back.
\section{Conclusions and Discussion}
In conclusion, we have shown that the superconducting stub geometry is
very different from the normal stub geometry in terms of its resonances
or anti-resonances. The presence of both electron and hole waves in the
stub leads to an unusual interference pattern, which causes resonances at a
non-unimodular value of the transmission. We perform a comprehensive
analysis of the analytic structure of the transmission amplitude for the
electron in the complex energy plane for the proposed geometry.
We have also discussed a possible application of the geometry in resonant
production of pure \scurrent by allowing incidence of spin polarised
electrons on the $\tstub$-stub geometry. At `resonance', we find that the
charge transport across the stub cancels out to zero, because normal
transmission and \car probability are both equal to $1/4$, whereas spin
transport is non-zero and becomes $1/2$ in the unit of spin that is
transmitted at resonance. It is also worth pointing out that this new
resonance is distinct from the standard resonances or anti-resonances at
$T=1$ or $T=0$ from the point of view of noise associated with these
resonances. As long as $T=1$ or $T=0$, the current fluctuations are
identically zero (\ie, we have zero noise) at zero
temperature~\cite{noiseButtiker}. But these new resonances allow for finite
noise at resonance even at zero temperature. Although our
analysis is strictly true for a single channel quantum wire, we expect the
new resonance to survive for a multi-channel wire, provided the inter
channel mixing is suppressed at the junction.

\bibliographystyle{eplbib}

\bibliography{supstubref}

\end{document}